\documentclass[12pt]{iopart}

\usepackage{graphicx}
\usepackage{siunitx}
\usepackage{float}

\usepackage{cite}
\usepackage{gensymb}
\usepackage{comment}

\usepackage[colorlinks=true, linkcolor=blue, citecolor=blue, urlcolor=blue]{hyperref}

\begin{document}

\title[]{Achieving precision in measuring birefringence characteristics of a periodically-poled Lithium Niobate waveguide}

\author{Stefan Kazmaier* and Kaisa Laiho}

\address{German Aerospace Center (DLR e.V.), Institute of Quantum Technologies, Wilhelm-Runge-Str. 10, 89081 Ulm, Germany\\ 
* Author to whom any correspondence should be addressed.}
\ead{stefan.kazmaier@dlr.de}
\vspace{10pt}

\begin{abstract}

The refraction of light in an optical medium is not only a subject of fundamental interest, but the refractive index also plays a crucial role in applications involving integrated and non-linear optics. One such application is photon-pair generation in waveguides with second-order nonlinearity. 
The spectral properties of the generated photon pairs are governed by the effective group refractive indices of the interacting modes, which normally are calculated via Sellmeier's equations for bulk crystals. However, in integrated optics, the effective group refractive indices experienced by the propagating modes can differ from those in bulk materials.
Therefore, we present an accurate, in-situ measurement technique for determining the birefringence characteristics of a structure with high reflective end facets by performing a Fourier transformation of the light transmission spectrum and apply this method to a periodically-poled Lithium Niobate waveguide resonator in the telecom wavelength range.
We directly predict important spectral figures of merit of the photon-pair generation process, which depend on the optical path length difference that can be resolved with a high precision of more than 16 standard deviations.
\end{abstract}

\section{Introduction}
The refractive index and its dispersion play an important role both in classical and quantum optics applications. For decades Sellmeier's equation has been used to model the refractive index and its dispersion in bulk material \cite{Sellmeier1872,Jundt:97}. However, in integrated optical devices the strong light-mode confinement can lead to deviations compared to the bulk media \cite{Kang:14,Roman-Rodriguez_2021}. To obtain accurate values for integrated optics, approximations or commercial solvers of Maxwell's equations may be exploited. However, even minor differences between the design and fabrication of such devices can lead to discrepancies in the final values. Moreover, Sellmeier's equation is not directly applicable at cryogenic temperatures making it challenging to deliver accurate predictions \cite{Thiele:20}.\\
One interesting process combining integrated and non-linear optics is the photon pair generation in waveguides (WGs) with second-order nonlinearity. 
A suitable material platform for integrated and non-linear optics is Lithium Niobate (LN), which has a moderately high optical second-order nonlinearity \cite{PhysRevA.111.013515}. Regarding quantum technologies such as quantum communication \cite{Ursin:07} or photonic quantum computing \cite{10.1063/1.5115814     } LN suits, for example, for generating photon pairs \cite{tanzilli_01} and entangled states \cite{PhysRevLett.82.2594}. Alternatively, LN can also be applied in the area of quantum simulation, such as to perform quantum walks \cite{Kruse_2013,Chapman_2024}.\\
In recent years non-linear optical WGs have shown their advantages over bulk crystals, due to their ease of handling, integrability and thus scalability and strong light-mode confinement, enabling the design of mode profiles and leading to higher brightness \cite{Wang_2020,Luo_2015}. The precise knowledge of the group indices and especially the birefringence in such structures is crucial for a number of reasons. While the refractive indices of the interacting modes dictate the conversion wavelength, the spectral form of the converted photons mostly stems from the group refractive indices. 
In other words, the group refractive indices strongly contribute to the frequency correlation between the photon pairs, which again determines their indistinguishability in the spectral degree of freedom. Although birefringence is usually required to achieve the desired operation of these devices, it leads to a temporal walk-off that often needs to be compensated accurately, especially in pulsed applications. Furthermore, the birefringence controls the bandwidth of the photon-pair emission in the case of narrow band pumping \cite{Luo_2015} and governs the cluster spacing in WGs embedded in resonators \cite{Brecht2016}.\\
In the past, several different approaches have been taken to access the relevant characteristics of the refractive index on several different non-linear optical platforms that are suitable for photon-pair generation. Linear optical methods include, for example, measurements with integrated Mach-Zehnder interferometers \cite{Dulkeith_2006} and Fabry-Pérot resonator (FPR) WGs \cite{Yoshizawa_1995,Laiho_2016}. Furthermore, non-linear optical methods that are based on fitting a refractive index model to the measurement of the spectral properties of either the generated photon pairs \cite{Misiaszek_2018} or the second-harmonic generation \cite{Neradovskiy_2021} have been employed, which on the downside increase the complexity of the required physical methods. For measuring the birefringence, one can also use a method involving polarization rotation in the WG, often called the "cross-polarizer" method, which can be very precise \cite{Rogers_2009,Heilmann_2014}, however, it cannot deliver the absolute measure of the group refractive index.\\
For the measurement of the effective group refractive index we implement the Fourier transform (FT) of the transmission spectrum of a FPR, introduced in 1997 by Hofstetter and Thorton \cite{Hofstetter:97}. It provides a versatile tool for accessing several linear optical parameters of the resonator devices. 
In early experiments it has been used for determining the optical losses and group indices in the FPRs used for semiconductor laser cavities \cite{720227}. However, this method has not only proven to be useful for measuring optical losses of non-linear optical materials \cite{Pergande:10,Thiel_2023}, but also for gaining information about the propagating spatial modes \cite{Pressl:15}, determining the effective group refractive indices \cite{Laiho_2016} and for extracting the spectral parameters of the photon-pair generation in waveguided structures surrounded by FPRs \cite{185054}.\\ 
Here, we extend this linear optical measurement procedure for accessing the birefringence of the group refractive index in an integrated optical device made of low refractive index material. 
We give a detailed insight into the measurement method and use it to resolve the birefringence of a periodically-poled LN WG (PPLN-WG) resonator with a high precision in situ measurement. We further show the advantage of utilizing the Fourier transform in extracting the birefringence, while accessing it directly from the resonator's free-spectral range seems to be too inaccurate and lie below the precision limit. 
Most importantly, we predict birefringence-based spectral parameters of the photon-pair generation process directly from the extracted optical path length difference for light modes polarized along the ordinary and extraordinary crystal axis in the telecommunication wavelength range. 
Our results show that in order to resolve the birefringence characteristics accurately enough, we require at least three meaning digits in the measurement of the optical path length difference stemming purely from the different group refractive indices.

\section{Theory}
A FPR is formed if light can travel back and forth in a transparent or even lossy medium between two flat reflecting facets placed at a distance. 
The destructive and constructive interference at the mirrors leads to intensity peaks and dips of the propagating light if expressed in terms of the wavelength. The transmission of the FPR depends on the reflectivity of the facets $R_\mathrm{j,1}$ and $R_\mathrm{j,2}$, the resonator length $L$ and the properties of the medium between the reflecting facets, such as its refractive index $n_\mathrm{j}(\omega_\mathrm{j})$ and optical loss $\alpha_\mathrm{j}$ with $j=e,o$ corresponding to the extraordinary and ordinary axis. The interference pattern of a classical FPR is described by an Airy function, whose amplitude $A_\mathrm{j}(\omega_\mathrm{j})$ is given by \cite{Brecht2016}
\begin{equation}
    A_\mathrm{j}(\omega_\mathrm{j}) = \frac{\sqrt{(1 - R_\mathrm{j,1})(1 - R_\mathrm{j,2})} e^{-\alpha_\mathrm{j} L / 2}}{1 - \sqrt{R_\mathrm{j,1} R_\mathrm{j,2}} e^{-\alpha_\mathrm{j} L} e^{i \phi_\mathrm{j}(\omega_\mathrm{j})}}
\end{equation}
with the roundtrip phase $\phi_\mathrm{j}(\omega_\mathrm{j})=2\omega_\mathrm{j} n_\mathrm{j}(\omega_\mathrm{j}) L/c$, $\omega=2\pi c/\lambda$ being the angular frequency of the propagating mode, the wavelength of light $\lambda$ and the speed of light $c$. The transmission spectrum of the resonator is proportional to $|A_\mathrm{j}(\omega_\mathrm{j})|^2$.\\ 
Following the treatment in \cite{Madsen:99}, we can express the free spectral range (FSR) of the transmission of the resonator in first-order approximation via $\omega_\mathrm{FSR}^\mathrm{j}=\frac{2\pi c}{Ln_\mathrm{g}^\mathrm{j}}$ in which $n_\mathrm{g}^\mathrm{j}$ is the group refractive index and the corresponding optical length $\mathrm{OL}^\mathrm{j}=n_\mathrm{g}^\mathrm{j}L$. The FSR is revealed by the transmission spectrum, from which the group index and its birefringence characteristics can be determined \cite{Pressl:15}. To achieve a high precision, the FT is employed for this purpose.
The spectra are recorded in terms of wavelength and converted to frequency. The FT returns the information in the time domain, where peaks appear approximately at the multiples of the round-trip time $\tau^\mathrm{j}$ of light in the resonator. Finally, $\tau^\mathrm{j}$ can be converted into optical length and by employing the resonator length into group velocity $v_\mathrm{g}^\mathrm{j}$ and group index by following
\begin{equation}
    n_\mathrm{g}^\mathrm{j}=\frac{c}{v_\mathrm{g}^\mathrm{j}}=c\frac{\tau^\mathrm{j}}{2L}. \label{eq:group-index}
\end{equation}
We compare our measurement results with the bulk model for the group index, available via Sellmeier's equation for 
a congruent PPLN bulk crystal with the Sellmeier coefficients for extraordinary and ordinary crystal axis found in \cite{Jundt:97}. To approximate the influence of the WG geometry commercial solvers of Maxwells equations or approximations like Marcatili's method for dielectric WGs can alternatively be used \cite{Marcatili_1969}. However, for simplicity, we chose the metallic WG approximation, which has been found useful in earlier studies \cite{Roman-Rodriguez_2021}. The approximation considers the surrounding edges of the WG as perfectly conducting, thus the electric field guided in the WG does not penetrate into the surrounding. By using this method one can approximate the refractive index for the extraordinary and ordinary axis of a rectangular waveguide simply as
\begin{equation}
    n^\mathrm{j}(\lambda, T, m_1, m_2, W, H) = n^\mathrm{j}(\lambda, T) + \left(\lambda \frac{ m_1 + 1}{2H}\right)^2 + \left(\lambda \frac{ m_2 + 1}{2W}\right)^2 \label{eq:metallic_approx}
\end{equation}
with the refractive index of the bulk material $n^\mathrm{j}(\lambda, T)$, from Sellmeier's equation, temperature $T$ and the WG height $H$ and width $W$. For the fundamental mode studied here $m_1=m_2=0$. In the calculation $H$ and $W$ take the same values as specified for our PPLN-WG. Finally, the group index is derived from the refractive index via 
$n^\mathrm{j}_\mathrm{g}(\lambda)=n^\mathrm{j}(\lambda)/\left(1+\frac{\lambda}{n^\mathrm{j}} \frac{\mathrm{d}n^\mathrm{j}(\lambda)}{\mathrm{d}\lambda}\right)$
for comparison of the results.
\section{Methods and Results}
\begin{figure}[b]
    \centering
    \includegraphics[width=0.4\textwidth]{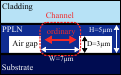}
    \caption{Cross-section of the PPLN-WG. The PPLN layer (blue) with the WG-channel (dashed red line illustrated as a guide for the eye) is confined by two air gaps (white) on the sides and two oxide layers on top and bottom. The PPLN layer is surrounded by the cladding (light blue) on top and the substrate (dark blue) on bottom. The ordinary crystal axis is shown with a red arrow.}
    \label{fig:Waveguide_skizze}
\end{figure}

\begin{figure}[t]
    \centering
    \includegraphics[width=0.55\textwidth]{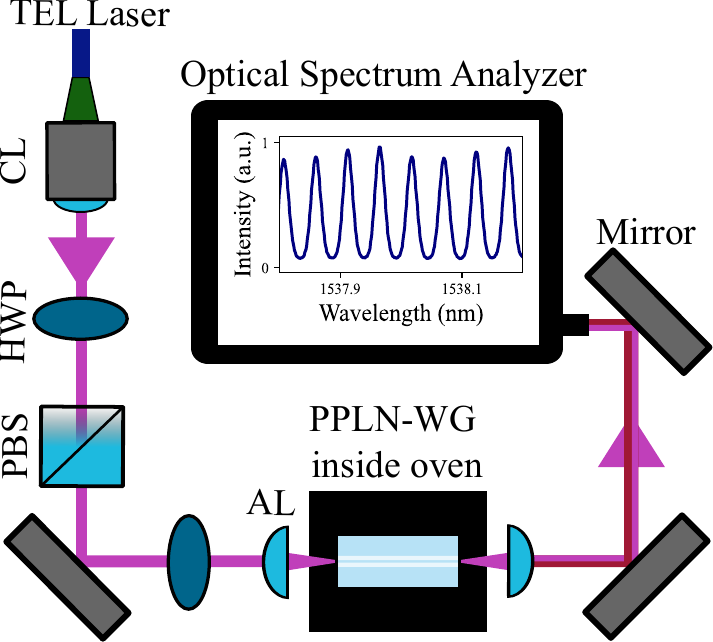}
    \caption{Schematic illustration of the experimental setup used to record the transmission spectrum of the PPLN-WG at telecommunication wavelengths. CL: Collimator, PBS: Polarizing Beam Splitter, HWP: Half-Wave Plate, AL: Aspheric Lens.}
    \label{fig:Setup_simplified_Paper_OL}
\end{figure}

\begin{figure}[t]
    \centering
    \includegraphics[width=0.6\textwidth]{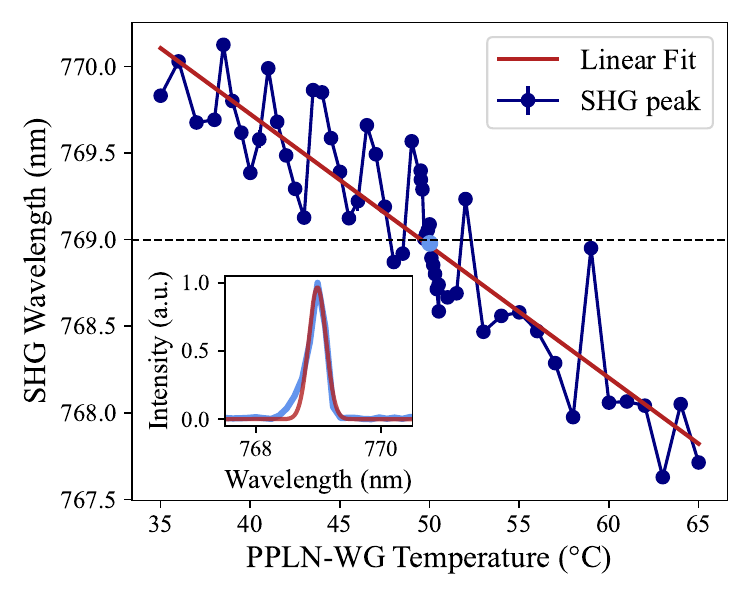}
    \caption{SHG wavelength as a function of the PPLN-WG temperature. The SHG peak positions are marked in blue with a linear fit in red. The inset shows one measured SHG spectrum corresponding to the SHG peak marked with light blue. The uncertainties in the SHG peaks are smaller than the used markers.}
    \label{fig:SHG_over_WL}
\end{figure}

\begin{figure}[t]
    \centering
    \includegraphics[width=\textwidth]{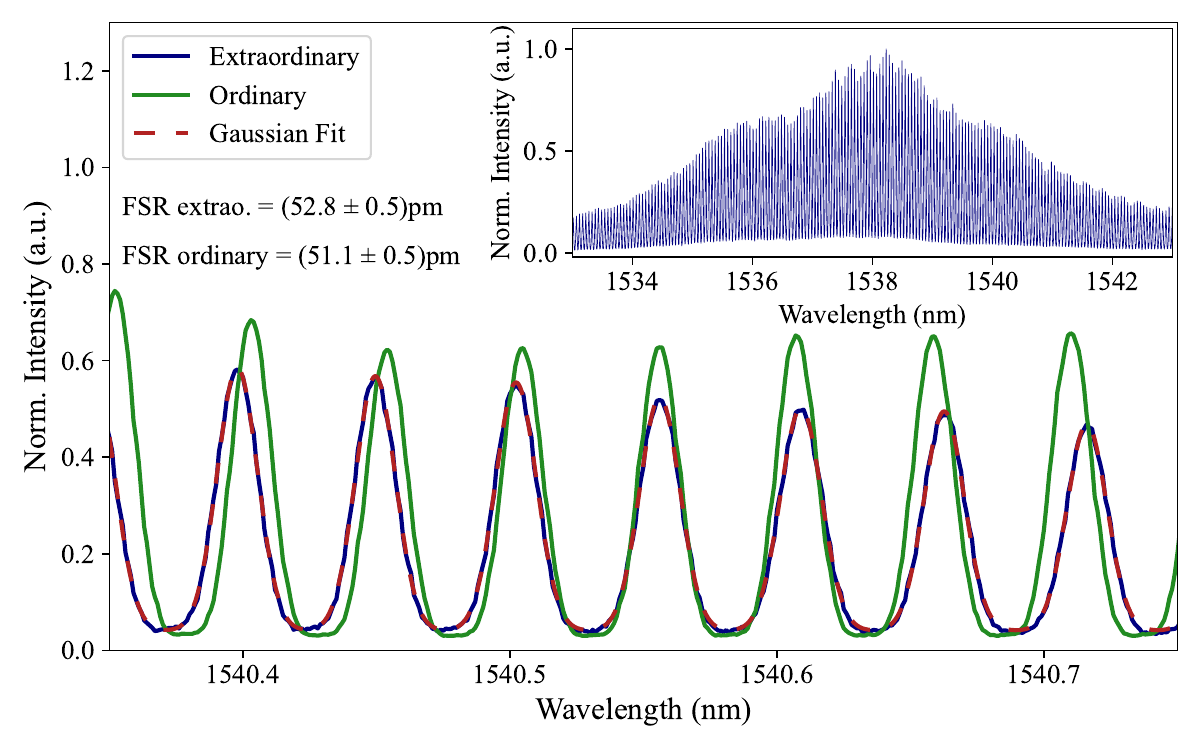}
    \caption{An individual transmission spectrum of the PPLN-WG resonator for the ordinary and extraordinary crystal axes, fitted with a sum of Gaussian distributions. In the inset the complete transmission spectrum is shown.}
    \label{fig:Transmission_Gauss_Fit_both_Pol}
\end{figure}

\begin{figure}[t]
    \centering
    \includegraphics[width=0.55\textwidth]{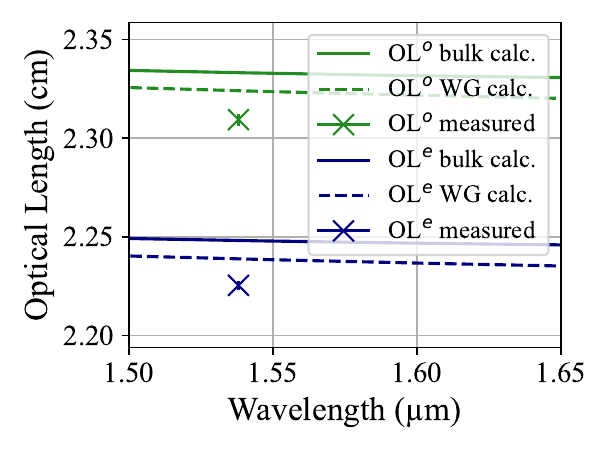}
    \caption{Optical lengths for ordinary and extraordinary crystal axes $\mathrm{OL}^\mathrm{o,e}$ in the PPLN-WG. The measured values are the weighted mean of nine transmission spectra each. The calculated values are determined for bulk PPLN via Sellmeier's equation and for the WG structure using the metallic approximation for a WG length of \SI{10.3}{mm}.}
    \label{fig:Group-Index_summary}
\end{figure}

In our experiments, we use a commercially available PPLN-WG with a length of $L=\SI{10.30\pm0.05}{\milli \meter}$, in which the error corresponds to the manufacturer's uncertainty for the specified measured length. It is made from congruent MgO:LN and manufactured by HCP. The cross-section of the device is illustrated in \fref{fig:Waveguide_skizze}.
Single-mode propagation in the WG channel is achieved by positioning air gaps on its sides and oxide layers both on top and bottom to the cladding and substrate layers. The ridge WG channel dimensions are $W=\SI{7}{\micro m}$ in width and $H=\SI{5}{\micro m}$ in height. Horizontal polarization corresponds to the ordinary (slow) crystal axes, while vertical polarization corresponds to the extraordinary (fast) crystal axes. The end facets of the PPLN-WG are coated with high reflection coatings $R_1=\SI{97\pm2}{\%}$ and $R_2=\SI{97.7\pm0.5}{\%}$, forming a FPR.\\
The optical measurement arrangement is shown in \fref{fig:Setup_simplified_Paper_OL}. 
We perform both second-harmonic generation (SHG) and linear optical transmission measurements in the telecommunication wavelength range. In the former case, we use a tunable continuous-wave telecom laser and in the latter case a tunable pulsed telecom laser with the repetition rate of \SI{40}{MHz}, pulse length of \SI{350}{fs}, and bandwidth of \SI{10}{nm}.
The collimated laser beam is directed through a half-wave plate (HWP) and polarizing beam splitter (PBS) to control the power and to set linear polarization. A second HWP is then used to adjust the polarization. The PPLN-WG is placed inside an oven to control its temperature for quasi-phase matching. 
The light coupling into and out of the PPLN-WG is achieved by using aspheric lenses. The light coupled out of the PPLN-WG is send either to a near infrared charge coupled device (CCD) spectrometer to measure SHG (not shown in \fref{fig:Setup_simplified_Paper_OL})  or to an optical spectrum analyzer (OSA) for measuring the transmission spectra at telecom wavelengths.\\
The periodic poling of the PPLN-WG is designed for type-II quasi-phase matching. It produces SHG from the fundamental wavelength of \SI{1538}{nm} to the second-harmonic wavelength of \SI{769}{nm} at around \SI{50}{\degree C}, as shown in \fref{fig:SHG_over_WL}, which defines our working environment.
The zigzag pattern of the SHG peaks most likely occurs due to a higher temperature dependent wavelength shift of the resonator, than the underlying phase-matching.
\\
The transmission spectrum of the WG cavity is measured from \SI{1533}{nm} to \SI{1543}{nm} for both the ordinary and extraordinary crystal axis, as shown in \fref{fig:Transmission_Gauss_Fit_both_Pol}. The pulsed laser is used, since a broad and continuous spectrum is beneficial for this measurement. The envelope in the inset in \fref{fig:Transmission_Gauss_Fit_both_Pol} is attributed to the non-Gaussian spectrum of the pump laser. Transmission spectra are recorded using an OSA with a resolution of \SI{20}{pm} and a step size of \SI{1}{pm}. 
The Airy profile expected from a FPR cavity is fitted using a sum of 190 Gaussian distributions, each variable in amplitude, width, and position. From the fit, the free spectral range (FSR) and full width at half maximum (FWHM) are determined by averaging over all individual peaks. The weighted average of nine measurements for both ordinary and extraordinary polarization is shown in \tref{tab:Cavity-Gauss-Fit}. The FSRs for ordinary and extraordinary polarization are expected to differ due to the birefringence of the PPLN-WG. The precision of the FSR measurement is sufficient to distinguish between both polarizations by $\Delta \text{FSR}=\SI{2.0\pm0.2}{pm}$, corresponding to \SI{10}{\ standard \ deviations \ (SD)}. Since the group index is the main influence on the FSR the dataset is further analyzed in Fourier space to extract the birefringence of the effective group refractive index. The resonator structure is convenient for this evaluation method, since it relies on a periodic transmission spectrum. \\
\begin{table}[t]
\caption{\label{tab:Cavity-Gauss-Fit} Summary of the PPLN-WG resonator investigation, including the weighted average FSR, FWHM and Finesse for the ordinary and extraordinary crystal axis for nine measurements each, with their weighted SD as uncertainty.}
\begin{indented}
\item[]\begin{tabular}{@{}llll}
\br
& FSR (pm)/(GHz) & FWHM (pm)/(GHz) & Finesse \\
\mr
extraordinary & \SI{52.9\pm0.1}{} / \SI{6.70\pm0.01}{}& \SI{18.10\pm0.08}{} / \SI{2.29\pm0.01}{} & \SI{2.92\pm0.02}{}  \\
ordinary & \SI{50.9\pm0.1}{} / \SI{6.45\pm0.01}{}& \SI{16.91\pm0.09}{} / \SI{2.14\pm0.01}{}& \SI{3.01\pm0.02}{}  \\
\br
\end{tabular}
\end{indented}
\end{table}
For this purpose, the same nine transmission spectra are used, covering a wavelength range of \SI{10}{nm} and a \SI{1}{pm} measurement step. To achieve high precision, a sufficiently large wavelength window is required in order to record a high number of peaks for the FT. We note that the wavelength window can become too large, because of group index dispersion, which in turn reduces the accuracy for large wavelength ranges. Thus, a balance between accuracy and precision is necessary. Here, the optimal range is found to be in a wavelength range of around \SI{10}{nm}, corresponding to approximately \SI{200}{peaks} of the FPR spectrum. The measurement step should be chosen as small as possible to maximize the resolution. \\ 
Each transmission spectrum is prepared for the FT by applying a Tukey window with the shape parameter \SI{0.25}{} to suppress any envelope functions. Although a narrower window, such as a Hanning window, has been suggested to improve the precision of the peaks in the FT, it did not improve the precision here. Probably because the window reduces the width of the transmission signal excessively. Finally, \SI{e5}{} zeros are added as zero padding to increase FT resolution, since it is the most important parameter here. The FT is performed using a Fast Fourier Transform (FFT) algorithm. Although a non-uniform Fourier transformation could be considered, since the transmission spectrum is measured over equal steps in wavelength, instead of wave number. Our analysis shows that the resulting error is at least an order of magnitude smaller than the given SD, since this mainly influences the peak shape and its influence grows for higher optical lengths \cite{Thiel_2023}. \\
\begin{figure}[t]
    \centering
    \includegraphics[width=\textwidth]{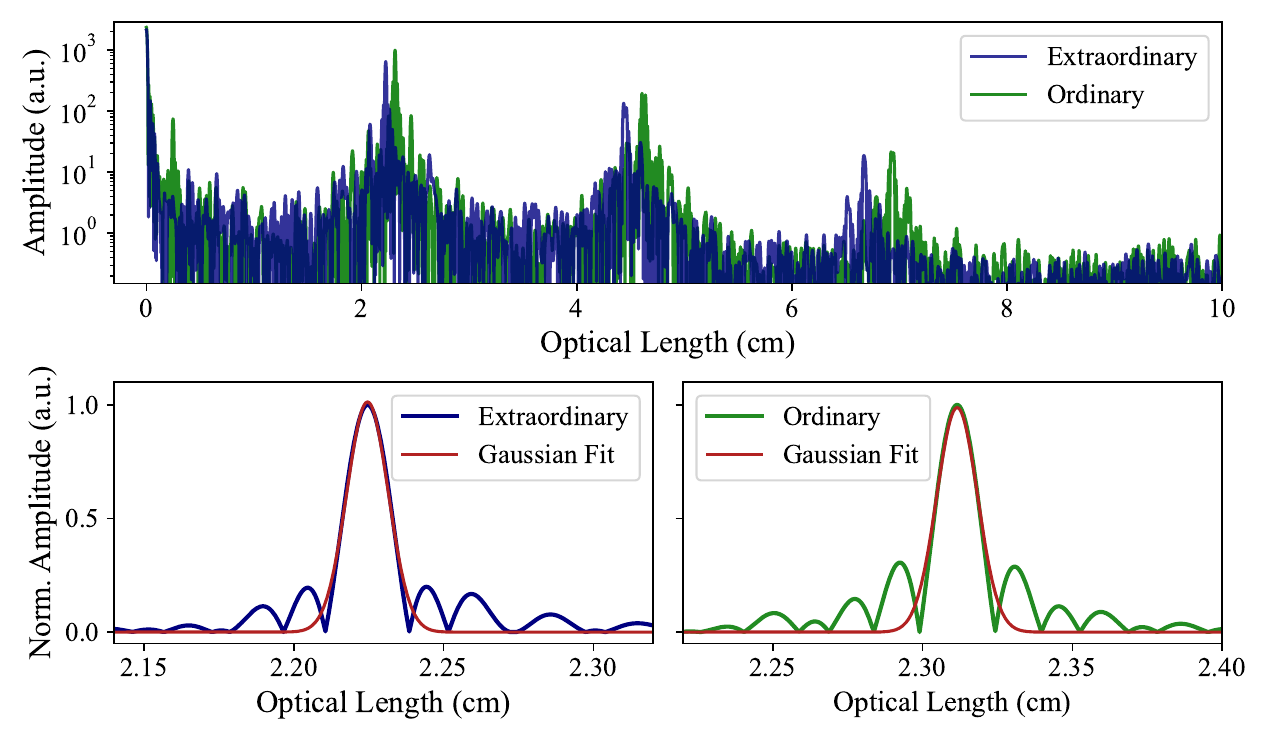}
    \caption{Optical length evaluation for the ordinary and extraordinary crystal axis. Top: FT for the first three peaks. Bottom: First peak of the FT with a Gaussian Fit to determine the optical length of the fundamental mode for the extraordinary crystal axis (left) and ordinary (right).}
    \label{fig:FFT_both_polarizations_complete_FFT_3_figures}
\end{figure}
In \fref{fig:FFT_both_polarizations_complete_FFT_3_figures} on the top, the first three peaks of the performed FT for ordinary and extraordinary polarization are plotted over the optical path length. The birefringence of the PPLN-WG is evident as a mismatch of the FT peaks, with twice the displacement for the second pair of peaks and three times for the third ones.
The horizontal axis of the FT is converted to optical length by using \eref{eq:group-index}. The position of the first peak in the FT gives the optical length experienced by the fundamental mode. A Gaussian distribution is fitted to determine the optical length $\mathrm{OL}^\mathrm{e,o} =n_\mathrm{g}^{e,o}L$ and the FWHM of the peak is used as uncertainty, as shown in \fref{fig:FFT_both_polarizations_complete_FFT_3_figures} on the bottom. The birefringence is clearly resolved in the first FT-peaks, with the difference increasing for peaks at multiple optical lengths providing consistency checks in order to validate our findings.\\
The extracted OLs are gained as a weighted median of all measurements together with the weighted standard deviation. In \fref{fig:Group-Index_summary} we present the results together with the theoretical calculations in which we apply \eref{eq:metallic_approx}.
The measured and calculated values of the group refractive indices for \SI{1538}{nm} at \SI{50}{\degree C} are given in \tref{tab:optical-length-values} together with the birefringence. The extracted group index difference is determined to be
$\Delta n_\mathrm{g}=n_\mathrm{g}^\mathrm{o}-n_\mathrm{g}^\mathrm{e}=\SI{0.081\pm0.006}{}$ including the inaccuracy of the WG length which also limits the uncertainty of the group indices to $n_\mathrm{g}^\mathrm{o}=\SI{2.242\pm0.014}{}$ and $n_\mathrm{g}^\mathrm{e}=\SI{2.160\pm0.012}{}$.
\begin{table}[t]
\begin{center}
\caption{\label{tab:optical-length-values} Summary of the optical length $\mathrm{OL}^\mathrm{e,o}$ and group index values $n_\mathrm{g}^\mathrm{e,o}$ for ordinary and extraordinary crystal axis and their birefringence $\Delta \mathrm{OL}$ and $\Delta n_\mathrm{g}$. Together with the measured values, we present the calculated values for the bulk PPLN from Sellmeier's equation and those from the metallic WG approximation with a WG length of \SI{10.3}{mm}. The difference between calculation and measurement $\Delta\mathrm{OL}^\mathrm{o,e}$ is given in SD.}
\resizebox{\textwidth}{!}{%
\begin{tabular}{@{}ccccccccc}
\br
& $\mathrm{OL}^\mathrm{o}$ & $\Delta \mathrm{OL}^\mathrm{o}$ & $\mathrm{OL}^\mathrm{e}$& $\Delta \mathrm{OL}^\mathrm{e}$ & $\Delta \mathrm{OL}$ & $n_\mathrm{g}^\mathrm{o}$  & $n_\mathrm{g}^\mathrm{e}$ & $\Delta n_\mathrm{g}$ \\
\mr
Bulk sim. & \SI{2.3333}{} & 8 & \SI{2.2482}{} & 11 &  0.0851 & \SI{2.2653}{} & \SI{2.1827}{} &  0.0826\\
WG sim. & \SI{2.3241}{} & 5 & \SI{2.2388}{} & 7 & 0.0853 & \SI{2.2564}{} & \SI{2.1736}{} & 0.0828\\
Measured & \SI{2.309\pm0.003}{} && \SI{2.225\pm0.002}{} && \SI{0.084\pm0.005}{} & \SI{2.242\pm0.014}{}& \SI{2.160\pm0.012}{} & \SI{0.081\pm0.006}{}\\
\br
\end{tabular}}
\end{center}
\end{table}
Finally, we predict key birefringence-based parameters that govern the photon-pair generation process in our PPLN-WG resonator and that only depend on the \emph{optical path length difference} $\Delta \mathrm{OL}=\Delta n_\mathrm{g}L=\SI{0.084\pm0.005}{cm}$, which is measured with a precision above \SI{16}{SD}. 
Therefore, they can directly be accessed just via the Fourier-transform analysis. We emphasize that although the FSR directly delivers information on the group index and its birefringence \cite{Puscas2011}, employing the FT analysis of the transmission spectra does reduce the uncertainty in the SD by a factor of \SI{1.6}{} here. First, the differential group delay $\Delta t=\Delta \mathrm{OL}/(2c)=\SI{1.4\pm0.1}{ps}$, which is known as the temporal walk-off of the generated photon wave packets.
Further, we extract two spectral parameters that are inversely proportional to the optical length difference.
In the limit of narrow band pump light the spectral bandwidth of the underlying phasematching envelope that corresponds to the photon-pair bandwidth, if the WG facets were uncoated, can be predicted via \cite{Luo_2015}
\begin{equation}
    \Delta \nu \approx \frac{5.566}{2\pi}\frac{c}{\Delta\mathrm{OL}}=\SI{320\pm20}{GHz},
\end{equation}
corresponding to a bandwidth of \SI{2.5\pm0.2}{nm} at the investigated wavelength of 1538nm.
Additionally, we predict the cluster spacing, which can be interpreted as the suppression of the photon-pair side peaks and which is caused by the interplay of pump, phasematching and the resonator transmission spectra. In the limit of narrow band pumping this spacing is given by \cite{Brecht2016} 
\begin{equation}
\Delta \nu_\mathrm{c}=\frac{c}{2\Delta\mathrm{OL}}=\SI{178\pm12}{GHz}
\end{equation}
corresponding to a suppression band of $2\Delta \nu_\mathrm{c}$ around the desired central peak, which delivers \SI{2.8\pm0.2}{nm} at the investigated wavelength of \SI{1538}{nm}.
These results deliver a valuable insight into the generation of photon pairs in the PPLN-WG. They indicate that a strong suppression of the spectral side-peaks over the phasematching envelope is possible in the limit of narrow band pumping.

\section{Discussion}
There are a number of different methods for measuring the group refractive index in non-linear optical integrated platforms. We use the FP-method since the facets of our waveguide have high reflective coatings at the telecommunication wavelengths. Further, the birefringence can be measured just in a linear transmission measurement and no changes to the existing experimental arrangement designed for the photon-pair generation are necessary, allowing for in situ measurements.\\ 
The precise knowledge about the behavior of the group refractive index is essential for simulating the properties of the photon-pair generation processes in non-linear optical devices. Further, it allows tailoring their properties to suit specific applications \cite{Kang:14}. Moreover, the accurate knowledge of the birefringence can enable the design of birefringence-free WGs in certain applications \cite{Schollhammer:17,Rogers_2009}.
For a correct simulation, the group refractive indices of the interacting modes should be known precisely. 
Given discrepancies with commonly used calculation methods, a simple measurement of the group refractive index becomes essential for a better understanding of the devices targeted for particular tasks. \\
For determining the group refractive indices from the measured optical lengths, the FP method requires no further length reference than the WG length itself. Indeed, the accuracy of the waveguide length measurement is one of the main restrictions in the accuracy of the extracted group index values. We further notice that the knowledge of the group index at the SHG wavelength can also provide important information on the spectral properties of photon pair generation process \cite{Pressl:15,Laiho_2016}. However, our PPLN-WG is coated with an anti-reflection coating around \SI{769}{nm} to achieve single-pass propagation of the pump pulse for the photon pair generation. Therefore, in this wavelength range we expect no resonator effects. Finally, we emphasize that with our method many key parameters of the studied device depending only on the birefringence in the telecommunication wavelength range can be predicted without any length reference.

\section{Conclusions}
We demonstrated a simple, in-situ, linear optical measurement procedure used for evaluating the birefringence in an optical media with highly reflectivity end facets via Fourier transformation. We measured the birefringence $\Delta n_\mathrm{g}=\SI{0.081\pm0.006}{}$ in a PPLN-WG resonator with a high precision and used it to predict key parameters of the photon pair generation process in the studied device that are based on birefringence, such as the differential group delay $\Delta t=\SI{1.4\pm0.1}{ps}$, the phasematching bandwidth $\Delta\nu=\SI{320\pm20}{GHz}$ and the cluster spacing $\Delta\nu_\mathrm{c}=\SI{178\pm12}{GHz}$. The device's operation temperature and wavelength were confirmed by detecting emission of the SHG, which is the reverse process of the photon-pair generation.
Finally, we investigated discrepancies to common calculation tools that are used for extracting the group indices revealing the necessity of an in-situ measurement for integrated-optics devices. Altogether, our results demonstrate that classical optical measurements provide a useful tool for the development of integrated non-linear optical devices targeted for the preparation of quantum light.

\section{Acknowledgements}
We thank Christoph Marquardt and the members of his group for their support with laboratory equipment.

\end{document}